\documentclass[longbibliography,floatfix,reprint,prb,aps,twocolumn,superscriptaddress]{revtex4-1}

\usepackage{bm}
\usepackage{ulem}
\usepackage{graphicx}
\usepackage{times}
\usepackage{soul}
\usepackage{color}
\usepackage{braket}
\usepackage{natbib}
\usepackage{floatrow}
\usepackage{physics}
\usepackage{comment}
\usepackage{xcolor}

\newcommand{\ba}{\begin{align}}
\newcommand{\ea}{\end{align}}

\newcommand{\kv}{\mathbf{k}}
\newcommand{\qv}{\mathbf{q}}

\newcommand{\EC}{\omega_\mathrm{C}}
\newcommand{\EX}{\omega_\mathrm{X}}

\newcommand{\EXXb}{\omega_\mathrm{XXB}}

\newcommand{\EL}{\omega_{\mathrm{LP}}}
\newcommand{\ELpump}{\omega_{\mathrm{LP},\sigma^+}}
\newcommand{\ELprobe}{\omega_{\mathrm{LP},\sigma^-}}
\newcommand{\EU}{\omega_{\mathrm{UP}}}

\newcommand{\tpump}{\Delta t_{\sigma^+}}
\newcommand{\tprobe}{\Delta t_{\sigma^-}}

\newcommand{\dbath}{n_\mathrm{b}}
\newcommand{\dbatht}{n_\mathrm{b,theo}}
\newcommand{\dimp}{n_\mathrm{i}}

\newcommand{\gib}{g_\mathrm{ib}}
\newcommand{\gii}{g_\mathrm{ii}}

\newcommand{\gXX}{\gamma_\mathrm{XX}}

\usepackage[colorlinks,
            citecolor = blue,
						linkcolor = blue,
						urlcolor  = blue,
						anchorcolor = blue,
						breaklinks = true
						]{hyperref}

\def \ETH{Institute for Quantum Electronics, ETH Z\"urich, CH-8093 Z\"urich, Switzerland}
\def \HD{Institute for Theoretical Physics, Heidelberg University, 69120 Heidelberg, Germany}
\def \MPQ{Max-Planck-Institute of Quantum Optics, 85748 Garching, Germany}
\begin{document}

\title{Bose polaron interactions in a cavity-coupled monolayer semiconductor}

\author{\surname{Li Bing Tan}}
\thanks{These two authors contributed equally}
\affiliation{\ETH}
\author{\surname{Oriana K. Diessel}}
\thanks{These two authors contributed equally}
\affiliation{\MPQ}
\author{\surname{Alexander Popert}}
\affiliation{\ETH}
\author{\surname{Richard Schmidt}}
\email{richard.schmidt@thphys.uni-heidelberg.de}
\affiliation{\HD}
\author{\surname{Atac Imamoglu}}
\email{imamoglu@phys.ethz.ch}
\affiliation{\ETH}
\author{\surname{Martin Kroner}}
\email{mkroner@phys.ethz.ch}
\affiliation{\ETH}

\begin{abstract}
{The interaction between a mobile quantum impurity and a bosonic bath leads to the formation of quasiparticles, termed Bose polarons. The elementary properties of Bose polarons, such as their mutual interactions, can differ drastically from those of the  bare impurities. Here, we explore Bose polaron physics in a two-dimensional nonequilibrium setting by injecting $\sigma^-$ polarised exciton-polariton impurities into a bath of coherent $\sigma^+$ polarised polaritons generated by resonant laser excitation of monolayer MoSe$_2$ embedded in an optical cavity. By exploiting a biexciton Feshbach resonance between the impurity and the bath polaritons, we tune the interacting system to the strong-coupling regime and demonstrate the coexistence of two new quasiparticle branches. Using time-resolved pump-probe measurements we observe how polaron dressing modifies the interaction between impurity polaritons. Remarkably, we find that the interactions between high-energy polaron quasiparticles, that are repulsive for small bath occupancy, can become attractive in the strong impurity-bath coupling regime. Our experiments provide the first direct measurement of Bose polaron-polaron interaction strength in any physical system and pave the way for exploration and control of many-body correlations in driven-dissipative settings.}

\end{abstract}

\maketitle

Interactions between quasiparticles play a key role across condensed matter physics, determining novel phases of matter and governing the quantum dynamics of many-body systems A paradigm for the physics of quasiparticles  are polarons arising from the interaction between a mobile impurity and a bath with a large number of degrees of freedom. The concept of polarons has been vital to develop effective descriptions of complex systems in terms of quasiparticles with properties such as an effective mass or lifetime that are modified compared to the bare impurity. Superconductivity provides a spectacular example for how polaron-polaron interactions could drastically deviate from those of the bare impurities: here, electronic quasiparticles form bound Cooper pairs due to bath mediated attraction even though bare impurities (the electrons) are subject to repulsive Coulomb interactions. Recent experimental investigation of single atomic impurities interacting with degenerate Fermi or Bose gases of ultracold atoms demonstrated the formation of Fermi or Bose polarons \cite{Schirotzek2009, Nascimbene2009, Kohstall2012, Koschorreck2012}. The impurity-bath interaction in ultracold atomic systems can be tuned using Feshbach resonances, allowing to enter the strong coupling regime where the impurity energy and lifetime is drastically altered \cite{Hu2016, Scazza2017,Yan_2020}. Recently, Fermi polarons were realised for the first time in the solid state physics where their formation explains the optical excitation spectrum of doped monolayer semiconductors \cite{MSidler2017}.

An outstanding challenge in both the atomic and solid state setting is to achieve a general understanding of how polaron quasiparticles interact with each other~\cite{Bruun2018b, Mistakidis_2019, Tan2020, Muir_2022}. Importantly, an open question remains as to whether interactions between polarons in these settings can fundamentally alter the character of quasiparticle interactions from repulsive to attractive as an emergent feature of the many-body character of the system. This question becomes of particular interest as it now becomes experimentally possible to enter the strong coupling regime where the interplay of exchange of low-energy excitations and the formation of significant dressing clouds can lead to strong non-perturbative effects.

In this letter we show that based on a mechanism of strong-coupling Bose polaron polariton formation, the interactions between fundamentally repulsive exciton-polariton quantum impurities in atomically thin semiconductors can not only be weakened but that the sign of interactions  be even reversed. Our observation demonstrates how Bose polaron formation can be exploited as a means to control interactions, opening avenues towards new unconventional mechanisms of induced pairing in van der Waals materials based on exciton or polariton exchange. Our observations are consistent with Bose polaron theory extended  to the description of a finite polaron density, where polaron formation acts as a mechanism to mediate attractive interactions. Remarkably we find that the dissipative nature of excitons plays only a subleading role in the theoretical predictions revealing the universal nature of the mechanism of exciton-mediated attraction.

\textbf{Experimental Scheme.} Our experiments are based on time-resolved pump-probe spectroscopy of exciton-polaritons created in a monolayer transition metal dichalcogenide that is strongly coupled to an optical microcavity. The pump generates a coherent state of lower exciton-polaritons of a large occupation number. These particles act as a quantum bath for the following probe pulse that generate a controllable density of exciton-polariton quantum impurities with a different valley quantum number. In absence of the bath these probe exciton-polaritons interact repulsively. 

Upon switching on interactions between the pump and probe exciton-polaritons using a biexciton Feshbach resonance, we observe that the dressing of the probe exciton-polaritons leads to the formation of repulsive (RP) and attractive (AP) Bose polarons  previously observed in ultracold atomic systems~\cite{Jorgensen_2016,Hu2016,Yan_2020}. In our solid-state setting we go  beyond these previous studies where impurities were extremely dilute, and enter the regime of finite polaron density. This reveals  that the dynamical dressing of exciton-polaritons leads not only to a smaller repulsive interactions between the higher energy RP quasiparticles, but ---by pushing the interspecies exciton interactions to the strong coupling limit--- the sign of interactions can be fully reversed.

\begin{figure*}[t!]
    \includegraphics[width=1\textwidth]{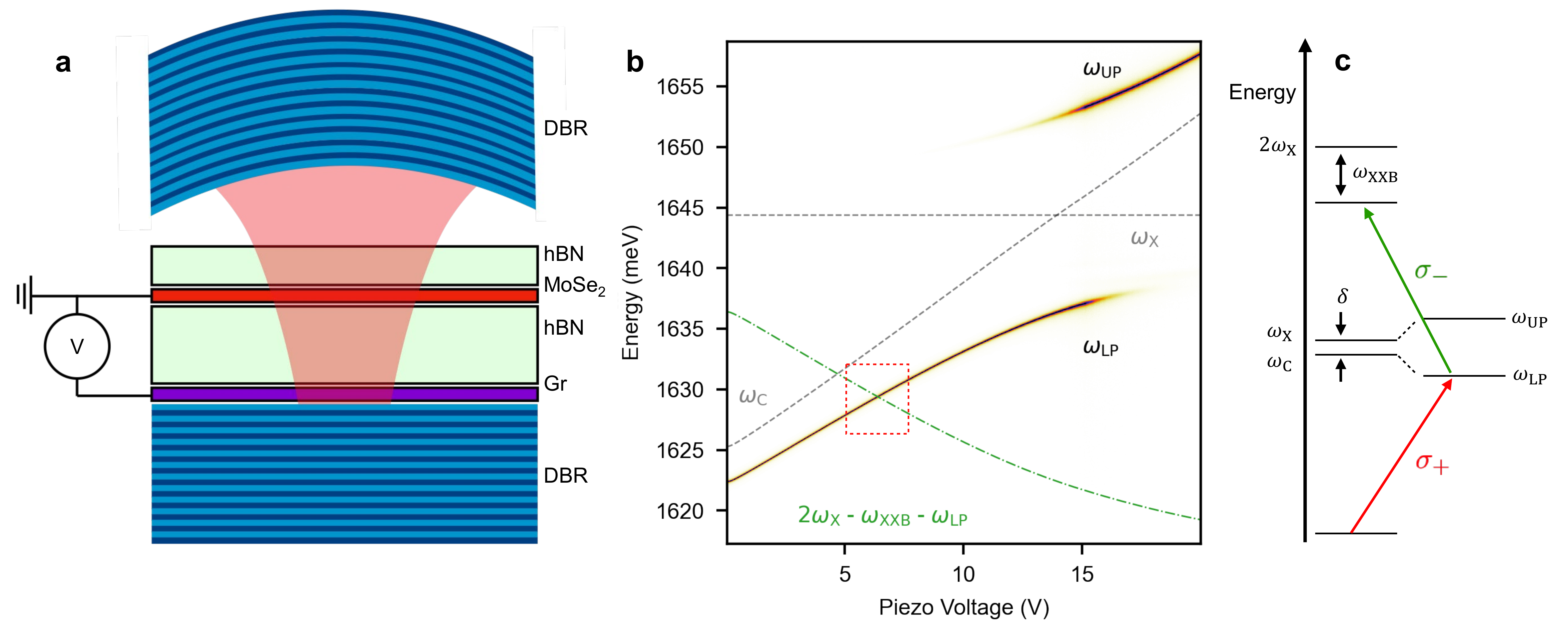}
    \caption{(a) Schematic of the device structure. The semiconductor cavity  consists of a van der Waals heterostructure deposited on a distributed Bragg reflector (DBR)- coated fused Silica substrate. The
    substrate along with a DBR-coated fibre with a concave facet on the core forms a 0D cavity. The heterostructure consists of a Molybdenum Diselenide (MoSe$_{2}$) monolayer (red) encapsulated by hexagonal Boron Nitride (hBN) layers (light green) and a graphene layer (purple) which acts as a gate. (b) The color plot shows the transmission spectrum of the exciton-polaritons as a function of piezo voltage controlling the cavity detuning $\delta=\EC-\EX$, measured with a broad band c.w. white light source. For illustration purposes, the maximum intensity of the spectrum is normalised to 1 for each piezo voltage. The bare exciton energy $\EX$ and the cavity energy $\EC$ are indicated by grey dashed lines. The biexciton resonance condition for a polariton in one valley in the presence of a polariton in the other is shown by the green dash-dotted line. The red dashed box indicates the detuning range around the resonance condition $2\EL=2\EX-\EXXb$, which is examined using pump-probe spectroscopy. (c) Energy level diagram to illustrate the experimental scheme. A K-valley polariton bath is resonantly created with a $\sigma^+$- polarised (pump) laser field (red arrow). The presence of this population allows for a resonant creation of a K'-valley polariton impurity population using a $\sigma^-$- polarised (probe) laser field (green arrow) to form a biexciton state.}
\label{fig: figure1}
\end{figure*}

Figure \ref{fig: figure1}(a) shows a sketch of the device we studied where a monolayer of Molybdenum Diselenide (MoSe$_{2}$) is embedded inside a zero-dimensional (0D) fibre cavity\cite{MSidler2017,Tan2020}. The MoSe$_2$ is encapsulated between two hexagonal Boron Nitride (hBN) layers, where the bottom layer serves as a dielectric spacer that separates the MoSe$_{2}$ from a graphene gate. It furthermore ensures that the graphene gate lies at the node, and the  MoSe$_{2}$ at the anti-node of the cavity electric field. The graphene and MoSe$_{2}$ are each contacted using metal electrodes made of Gold and Titanium. A gate voltage $V$ is applied to operate the device in the charge neutrality regime of the MoSe$_2$ monolayer. The heterostructure is deposited on a fused Silica substrate whose surface is coated with a distributed Bragg reflector (DBR) mirror. The cavity is completed by a DBR-coated optical fibre facet with a concave dimple fabricated on the core. (see Methods for details). 

The elementary excitations of this coupled system are exciton-polaritons, whose properties we investigate at liquid Helium temperatures \cite{MSidler2017,Tan2020}. In Fig.~\ref{fig: figure1}(b) we show the normalised transmission spectrum of the upper  and lower  exciton-polaritons of energy $\EU$ and $\EL$, respectively, as a function of the voltage applied to the piezoelectric positioner that controls the cavity length and thereby the energy of the cavity mode  $\EC$. The two excitonic species belonging to the K and K' valleys can be individually addressed by $\sigma^-$- and $\sigma^+$- polarised light. They are degenerate in the absence of a magnetic field and hence exhibit the same transmission spectrum when probed by a low intensity continuous wave (c.w.) white light source. We exclusively study excitations of the lower branch in our experiments. Consequently, we will use the term polariton to denote lower exciton-polaritons.

To investigate Bose polaron physics, we exploit the existence of a polariton Feshbach resonance between polaritons with opposite chirality, previously demonstrated in GaAs quantum wells embedded in a monolithic 2D DBR cavity \cite{Takemura2014}. In contrast to this prior experiment we use a 0D cavity.
However, despite the 0D nature of the photon, excitons are mobile in the 2D monolayer so that the excition-polaritons do acquire an effective 2D dispersion relation and thus a model for mobile quantum impurities (and a mobile bath) applies~\cite{Carusotto2013}.

We resonantly excite the system with a narrow band ($\leq 0.5$ meV, corresponding to $\tpump \sim 3$~ps pulse length), strong $\sigma^{+}$-polarised pump pulse creating a bath of exciton-polaritons in the K-valley (of density $\dbath$). We then measure the transmission spectrum of the K' impurity  exciton-polaritons (of density $\dimp$) with a weak $\sigma^{-}$-polarised $7$ meV broad-band pulse $(\tprobe \sim 200$~fs). The time delay $\tau$ between the pump pulse and the probe pulse, defined as time difference between the arrival time of the peak of the probe pulse and the pump pulse, can be varied in our experiment (see Methods).
Fig.~\ref{fig: figure1}(c) shows the relevant energy states addressed in our experiment. In MoSe$_{2}$, two opposite spin/valley/polarisation exciton-polaritons can form a biexciton bound state of binding energy $\EXXb$ below the two-exciton threshold at $2\EX$  \cite{Hao2017}. The green dash-dotted line in Fig. \ref{fig: figure1} (b) shows the resonance condition $\ELpump + \ELprobe = 2 \EX -\EXXb$ between the pump (bath) and probe (impurity) polariton energies with the biexciton energy as a function of the cavity detuning. Here we use $\EXXb = 29$~meV, obtained by  recent ac-Stark shift measurements on bare MoSe$_2$ excitons~\cite{Uto2022}\,\footnote{This observation implies a biexciton binding energy larger than what has been reported in earlier experiments which yielded $\EXXb = 20$ meV~\cite{Hao2017,Yong2018}, and is not consistent with our findings.}.

\begin{figure*}[t!]
            \includegraphics[width=1\textwidth]{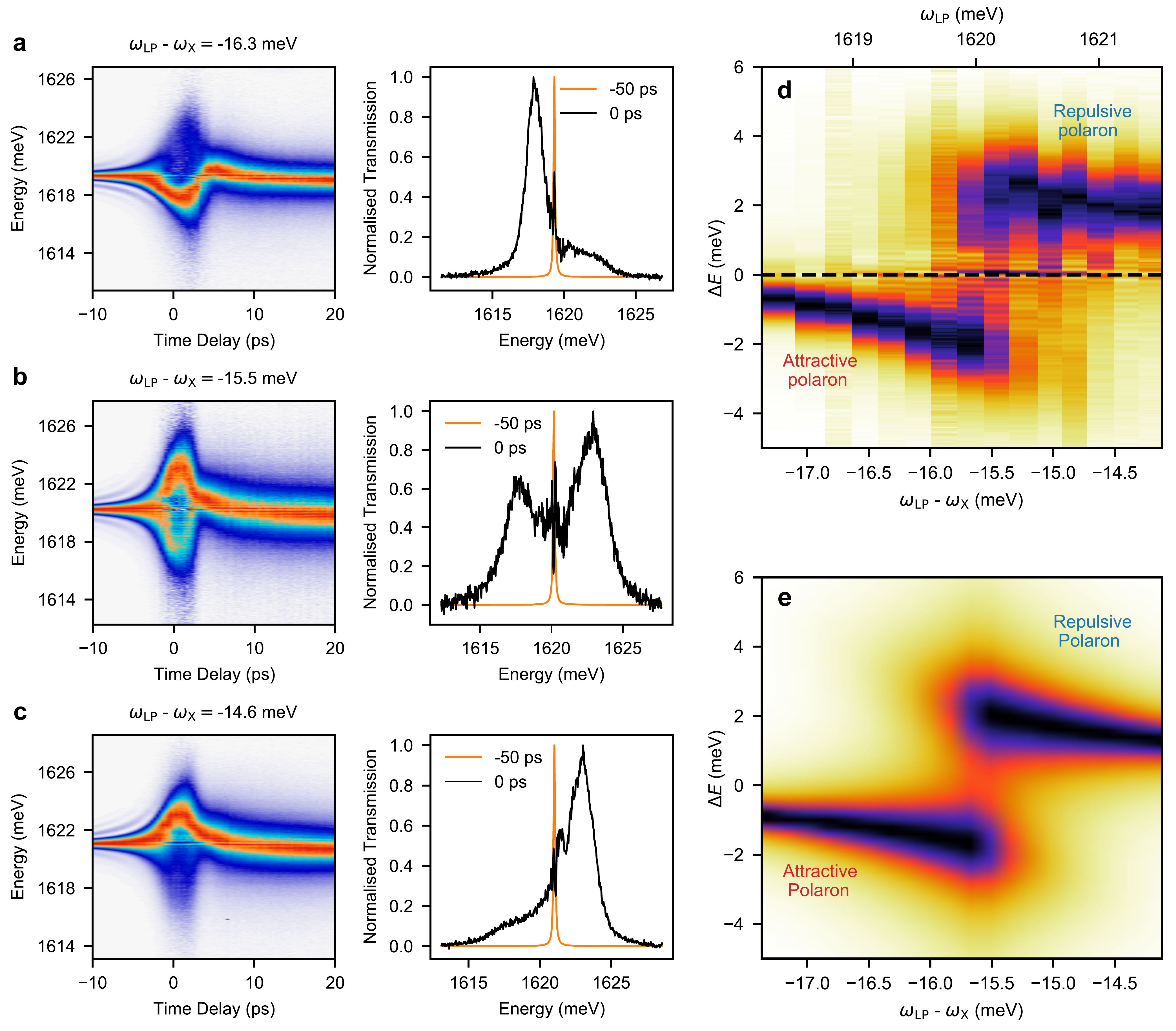}
    \caption{(a) - (c) Left panel: Normalised probe transmission as a function of time delay for $\omega_{\text{LP}} - \omega_{\text{X}} = -16.3$meV, $-15.5$meV and $-14.6$meV at an estimated bath exciton-polariton density of $n_{\text{b}} = 6.3 \times 10^{11}$cm$^{-2}$ at zero time delay, respectively. At negative time delays, i.e. before the arrival of the pump pulse which creates the bath polaritons, the transmitted probe spectrum peaks at the lower exciton-polariton energy for all detunings. Starting from $-2$ ps towards positive time delays, the probe population has significant temporal overlap with the pumped population which leads to the formation of Bose polarons. At resonance where $\omega_{\text{LP}} - \omega_{\text{X}} \approx -15.5$meV, the attractive and repulsive branches have approximately equal weight. At red (blue) detuning, the attractive (repulsive) branch dominates. (a) - (c) Right panel: Linecuts of the normalised probe transmission spectrum at negative (orange) and zero (black) time delays. (d) Transmission spectrum as measured from the undressed exciton-polariton energy as a function of detuning at zero time delay exhibiting the attractive and repulsive polaron branches. At zero time delay, the bath density introduced by the pump pulse is estimated to be $n_{\text{b}} = 6.3 \times 10^{11}$ cm$^{-2}$. (e) Simulated Bose polaron spectrum with the fit parameters $n_{\text{b,theo}} = 2.5 \times 10^{11} \text{ cm}^{-2}$ and $\gamma_{\text{XX}} = 2.5$ \text{ meV}.}
\label{fig: figure2}
\end{figure*}

\textbf{Bose polaron formation.} The left panel of Figure \ref{fig: figure2} (a)-(c) shows the normalised  probe transmission spectrum as a function of time delay when the polariton is tuned to $16.3, 15.5$ and $14.6$ meV below the exciton energy, respectively. At negative time delays $\tau < -2$ ps, the probe population is generated in the sample before the bath population. As expected,  the transmission remains  peaked at the polariton energy. Starting from $\tau = -2$ ps, the temporal overlap between  bath and probe population becomes significant and increases towards positive time delays. It is in this regime that we observe a clear modification of the transmission spectrum. At fixed detuning $\EL - \omega_{\text{X}} = -16.3$ meV (Fig.~\ref{fig: figure2} (a)), the peak of the transmission spectrum begins to redshift and its magnitude becomes largest at around $\tau = 2$ ps before relaxing back to the initial polariton energy. When the sum of the energy of bath and impurity polaritons equals $2\EX-\EXXb$ (Fig.~\ref{fig: figure2} (b)), the transmission splits into two distinct peaks at zero delay which persists until around $\tau = 4$ ps. In contrast, for $\EL - \omega_{\text{X}} = -14.6$ meV (Fig.~\ref{fig: figure2} (c)), the peak of the transmission blueshifts for $|\tau| \le 4$~ps. 

Figure \ref{fig: figure2} (d) shows the normalised transmission spectrum (at $\tau=0$) obtained as a function of detuning and expressed with respect to the energy of the polariton impurity  in absence of the bath (at $\tau = -50$ ps). The peak bath density is determined to be $\dbath = 6.3 \times 10^{11}$ cm$^{-2}$ (See Methods). We  observe that for  two-polariton energies ($\ELpump + \ELprobe$) below (above) the biexciton energy, the probe polariton resonance is shifted to lower (higher) energies, and as the energy $\EL - \omega_{\text{X}} \sim -15.5$ meV is approached the two branches  exhibit an avoided crossing.

\begin{figure*}[t!]
            \includegraphics[width=\textwidth]{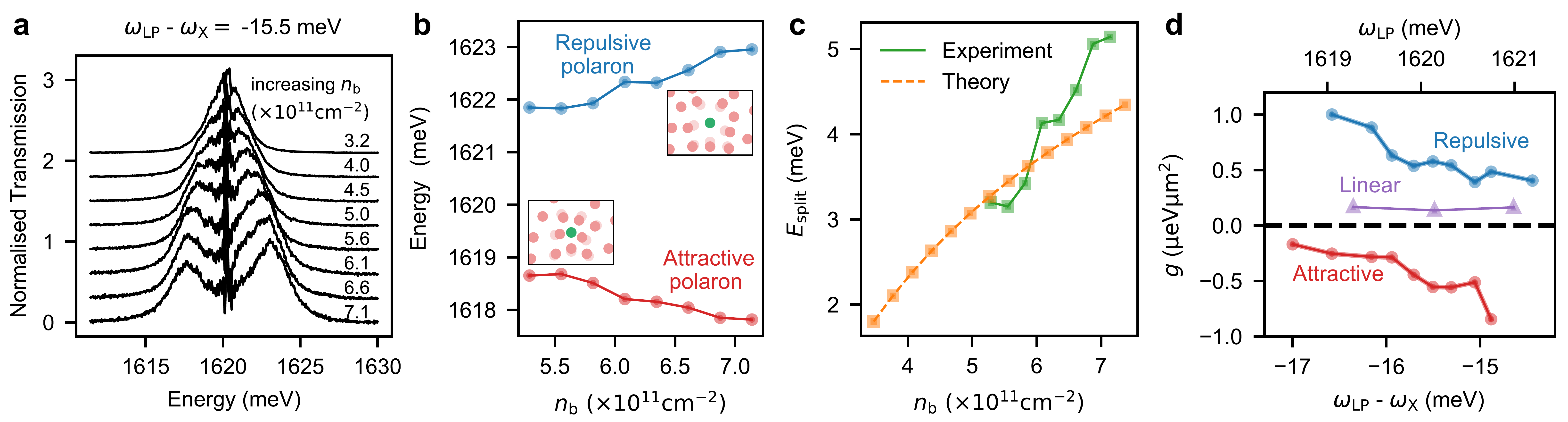}
    \caption{(a) Normalised Bose polaron  transmission spectrum for bath densities $n_{\text{b}} = 3.2\times 10^{11}$cm$^{-2}$ to $7.1 \times 10^{11}$cm$^{-2}$ at $\tau = 0$ ps. At low densities, the splitting between the two branches is much smaller than their linewidths. As the bath density is increased, the attractive and repulsive branches become clearly distinguishable as the splitting grows. The spectra for different bath densities are offset for clarity. (b) The peak energies of the repulsive (attractive) branch in blue (red) circles as a function of bath density. The energies are obtained from Lorentzian fits to the measured spectrum. (c) Comparison of the measured (green) and calculated (orange) energy splittings between the attractive and the repulsive branch as a function of bath density. (d) Impurity-bath interaction strength $\gib$ of the repulsive (attractive) branch in blue (red) circles as a function of detuning. The purple triangles correspond to the polariton-polariton interaction strength ($\gii(\dbath=0)$) measured in linear polarisation as a function of the detuning.}
\label{fig: figure3}
\end{figure*}

To explore the physics behind our observation, we follow the theoretical framework developed in Ref.~\cite{Levinsen2019} to model the system by the Hamiltonian
\begin{align}
\mathcal{H}=\sum_{\kv}
&\Big[
\omega_{\text{X}}(\kv) b_{\kv}^{\dagger}b_{\kv}^{\phantom{\dagger}}
+\omega_{\text{C}}(\kv) c_{\kv}^{\dagger}c_{\kv}^{\phantom{\dagger}}
+\frac{\Omega}{2}\left(b_{\kv}^{\dagger}c_{\kv}^{\phantom{\dagger}}+\text{h.c.}\right)
\nonumber
\\
&+\left(\omega_\mathrm{LP}(\kv)-\omega_\mathrm{LP}(0)\right) L_{\kv}^{\dagger}L_{\kv}^{\phantom{\dagger}}
\Big]
\nonumber
\\
&+
\frac{1}{V}\sum_{\kv,\kv',\qv}g^{\phantom{\dagger}}_\text{{ib}}(\kv,\kv')L^{\dagger}_{\kv}b^{\dagger}_{\qv-\kv}b^{\phantom{\dagger}}_{\qv-\kv'}L^{\phantom{\dagger}}_{\kv'}. 
\label{Eq:Hamiltonian}
\end{align}
We recognise that due to their large binding energy in TMD, excitons can be regarded as pointlike particles at the densities of our experiment. 
Here the probe excitons ($b^{\dagger}_\kv$) and photons ($c^{\dagger}_\kv$) with their respective energies $\omega_\mathrm{X}(\kv)=\omega_\mathrm{X}+\kv^2/2m_\mathrm{X}$ and $\omega_\mathrm{C}(\kv)=\omega_\mathrm{X}+\delta+\kv^2/2m_\mathrm{C}$ are coupled by the light-matter interaction $\Omega$. Because the pump laser is tuned to the lower polariton resonance, the rotation into the polariton basis (with respective operators $L^{\dagger}_{\mathbf{k}}$ and energies $\omega_{\text{LP}}(\kv)$) has already been performed for the bath excitations. The interaction between pump and probe excitons is modeled by a contact interaction of strength $g$, which in polariton basis results in the interaction $\gib^{\phantom{\dagger}}\!(\kv,\kv')=g\,\text{cos}\,\theta_{\kv}\text{cos}\,\theta_{\kv'}$ between pump polaritons and probe excitons. Here the Hopfield factors   $\text{cos}\,\theta_{\kv}$ correspond to the exciton fraction in the pump polariton at momentum $\kv$.
The finite lifetimes of probe photons and pump polaritons are included in the model as phenomenological parameters (see Methods).

The modified polariton impurity eigenstates can be approximately described using the Ansatz~\cite{Levinsen2019, Bastarrachea-Magnani2019}
\begin{align}
    |\psi\rangle&=\left(
    \phi^\mathrm{C} c_\mathbf{0}^{\dagger}+\phi^\mathrm{X} b_{\mathbf{0}}^{\dagger}+\frac{1}{\sqrt{NV}}\sum_{\kv}\phi_{\kv}^{\phantom{\dagger}}b_{-\kv}^{\dagger}L_{\kv}^{\dagger}L_{\mathbf{0}}^{\phantom{\dagger}}  \right)|\text{L}\rangle\nonumber\\
&\equiv\hat a^{\dagger}|\text{L}\rangle
\label{Eq:Wavefunction}
\end{align}
where $\phi^\mathrm{C},\,\phi^\mathrm{X}$, and $\phi_{\kv}$ are variational parameters.
The resonant pump laser creates a coherent state of lower polaritons, which we approximate by the Fock state $|\text{L}\rangle=(N!)^{-1/2}(L^{\dagger}_{\mathbf{0}})^N|\text{0}\rangle$, which is an accurate approximation for a coherent state in the limit of large $N$. The first two terms account for the hybridisation of the photon and the exciton into polaritons, while the third term reflects the dressing of the exciton by bath polaritons, leading to the formation of a polaron cloud. 

Using the Ansatz~\eqref{Eq:Wavefunction}, we calculate the transmission probe spectrum (Fig.~\ref{fig: figure2}(e)), using the independently measured biexciton binding energy $\EXXb$ as input. The biexciton decay width $\gXX$ in turn is treated as a fit parameter due to the absence of an experimentally determined value. More importantly, we also take the peak bath density $\dbatht$ as an additional fit parameter. Using this procedure we find a theoretical value $\dbatht$ that is smaller than the experimentally estimated value by a factor $\beta = 2.6$; this difference can be attributed to the fact that  the absolute value of the experimentally estimated polariton density $\dbath$ is subject to large uncertainty. Using only the two fit parameters $\dbatht$ and $\gXX$ we find remarkable agreement between theory and experiment across the full range of detunings and frequencies. This agreement gives a solid foundation
to the interpretation of the observed response of the probe pulse as being due to the formation of APs and RPs, arising from the dynamical dressing of the polariton impurity by the polariton bath. 

While Bose polarons exhibiting RP and AP branches have previously been observed in cold atom systems~\cite{Jorgensen_2016,Hu2016,Yan_2020}, our transmission data is the first evidence for their existence in 2D materials. To further characterise their properties and the role of medium dressing in this light-matter coupled system, we study the polaron spectrum measured at zero time delay and at fixed detuning $\omega_{\text{LP}} - \omega_{\text{X}} = -15.5$ meV for various bath densities, see Fig.~\ref{fig: figure3} (a). At low densities, there is no discernible splitting between the two branches. As the bath density is increased, the finite splitting appears and grows, rendering the two peaks distinct. By fitting each spectrum with a sum of two Lorentzians, we extract the peak energies of each polaron branch (Fig.~\ref{fig: figure3} (b)) and their corresponding splitting $E_\mathrm{split}$ (Fig.~\ref{fig: figure3} (c)) as a function of the bath density. 

Both, the RP and AP states and  their respective energies are captured by wave functions of the form in Eq.~\eqref{Eq:Wavefunction}. The predicted splitting of these are shown as orange squares in Fig.~\ref{fig: figure3} (c). In this calculation the theoretical densities are rescaled by the fixed factor $\beta=2.6$, obtained from the measurement of the transmission spectrum in Fig.~\ref{fig: figure2}. 
We find that the theoretical model predicts a slight shift with density for the detuning at which the ``jump'' in the spectral weight from  the attractive to the repulsive branch occurs, which was also predicted for cold atomic gases. This shift is not observed in the experiment which may be attributed to finite range effects (that are to a large degree negligible in the cold atom setting) and not included in our theoretical model. Accounting for this we extract the predicted energy splitting $E_{\text{split}}$ for the detuning at which the transfer of weight occurs in the theoretical data. 
The theory is in good qualitative agreement with the experimental data.
However, while we experimentally observe a linear scaling of the splitting with respect to the bath density $\dbath$, the scaling in theory is sublinear. A possible source of the different scaling might again arise from finite range corrections.

For sufficiently large bath density we can approximate the impurity-bath interaction as a Kerr type nonlinearity, and obtain an \textit{effective} interaction strength $\gib^{\text{eff}}$ between the impurity and bath polaritons from the gradient of the peak polaron energies with respect to the bath density: $\Delta \omega = \gib^{\text{eff}}\dbath\dimp$. As shown in Fig.~\ref{fig: figure3} (d), by tuning across the biexciton resonance, the interaction strength varies from $\gib^{\text{eff}} = 0.4 \mu$eV $\mu$m$^{2}$ to $1 \mu$eV $\mu$m$^{2}$ on the RP branch and from $\gib^{\text{eff}} = -0.2 \mu$eV $\mu$m$^{2}$ to $-0.8 \mu$eV $\mu$m$^{2}$ on the AP branch. This finding implies that the magnitude of the interaction strength between opposite-spin polaritons can be enhanced by a factor of up to $\sim 5$ times in comparison to that of parallel-spin polaritons in the absence of a bath ($\gii(\dbath=0) = 0.2 \mu$eV $\mu$m$^{2}$). Most importantly, not only the magnitude but also the sign of $\gib$ is tunable. This can be understood as a consequence of the polariton-biexciton Feshbach resonance where by tuning the energy of a scattering state across a bound state, interactions can be enhanced and their sign can be reversed \cite{Wouters2007, Carusotto2010, Takemura2014}.

\begin{figure}[t!]
            \includegraphics[width=\textwidth]{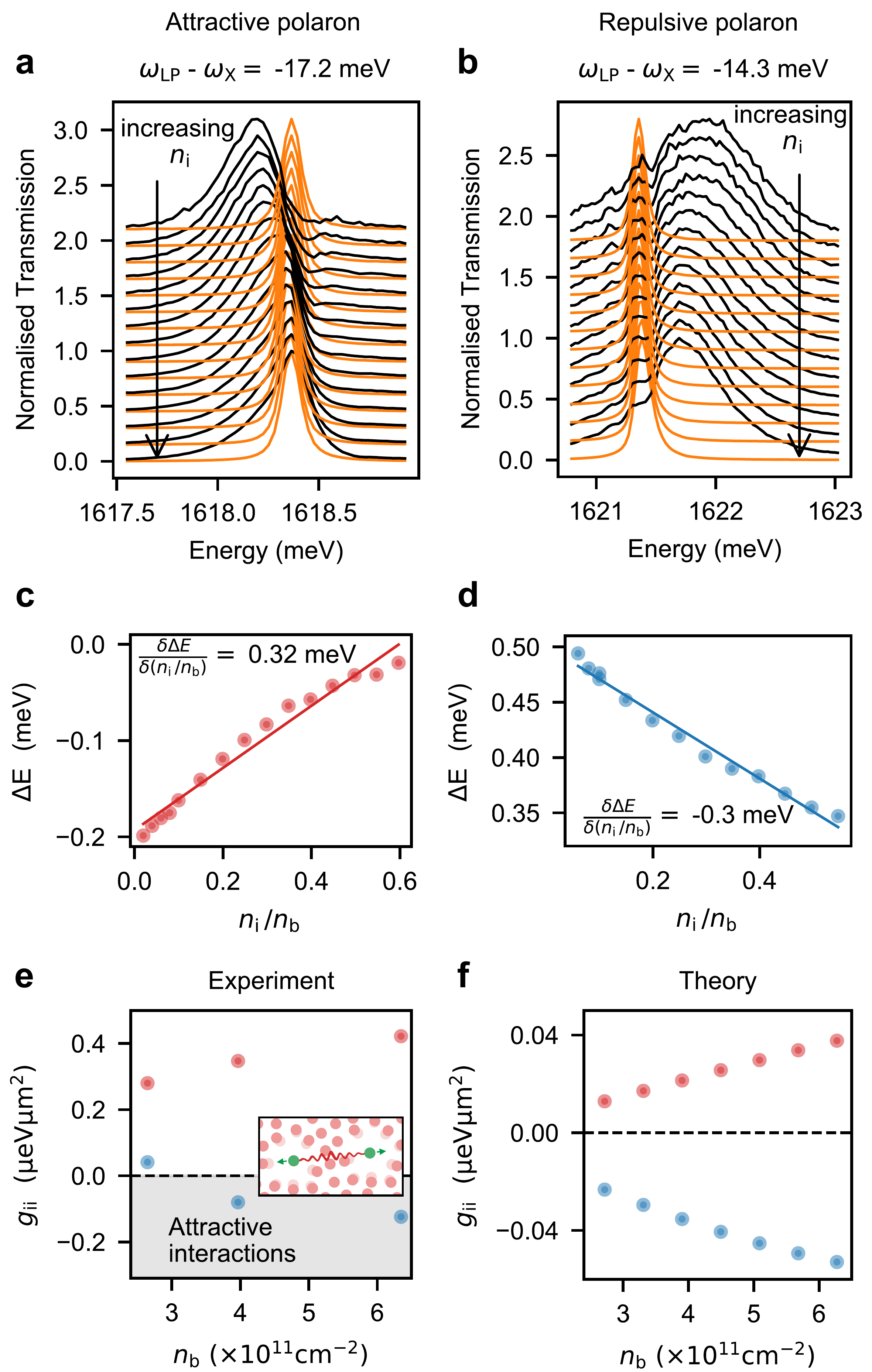}
    \caption{Top panel: Normalised impurity density dependent polaron ((a) attractive and (b) repulsive) spectrum in black lines ($\tau = 0$ ps) and the corresponding undressed exciton-polariton spectrum (measured at $\tau=-50$ ps) in orange lines. The bath density was $\dbath = 2.6 \times 10^{11}$ cm$^{-2}$. The spectra for different impurity density are plotted with an offset for clarity . Middle panel: Impurity-impurity interaction shift of the repulsive (c) and attractive (d) branch as a function of the ratio between impurity and bath density, at $\dbath = 2.6 \times 10^{11}$ cm$^{-2}$. Bottom panel: Experimentally (e) and theoretically (f) obtained impurity-impurity interactions strengths as a function of the bath density. In both cases the values for the attractive (repulsive) polaron are given by the red (blue) dots. The interaction strength has been measured with respect to the undressed polariton spectrum which included polariton-polariton interactions in the same valley, implying that the interactions strength values above the gray shaded area are repulsive. The inset in (e) symbolises the repulsive interactions between impurities (green dots) repulsively dressed by excitations of bath particles (red dots).}
    \label{fig: figure4}
\end{figure}

\textbf{Induced interactions between Bose polarons.} Next, we address the question of how the interaction between two impurities  are modified when they form polaronic quasiparticles due their coupling to a bosonic bath. To this end, we now keep the bath density fixed at $\dbath = 2.6 \times10^{11}$ cm$^{-2}$ and monitor the changes in the energy of the AP and RP excitations as a function of the impurity density ranging from $\dimp=0.15\times10^{11}$ cm$^{-2}$ to $1.4\times10^{11}$ cm$^{-2}$. 
In the top panel of Figure \ref{fig: figure4}, the polaron spectrum measured at zero time delay is shown as black lines for various impurity densities.  The probe spectrum, measured at $\tau = -50$~ps where the impurity is undressed (orange lines in Fig.~\ref{fig: figure4}(a) and (b)), was recorded for each probe power and serves as a control experiment. In these spectra the residual energy shifts arise due to interactions between the impurities in the presence of the long-lived incoherent heating effects induced by the previous pump pulses (see Fig. \ref{fig:EDF}).

For the AP-polariton branch we find that the energy of the dressed impurities blueshifts as their density is increased. For RP-polaritons on the other hand, we find the opposite effect: increasing the impurity density leads to a redshift. That means increasing the density of repulsive polarons lowers the energy to create further repulsive polarons. By taking the gradient of the energy shifts with respect to the impurity density, we can extract the effective interaction strength between the polaron-polaritons (see Figures \ref{fig: figure4} (c) and (d)). For APs this leads to a repulsive interaction strength $\gii = 0.12 \mu$eV $\mu$m$^{2}$ while RP-polaritons experience an attractive interaction strength of  $\gii = -0.11 \mu$eV $\mu$m$^{2}$. These interaction values have been measured relative to the polariton-polariton interaction strength without bath interactions $\gii(\dbath=0)=0.2 \mu$eV $\mu$m$^{2}$ (see purple data in Figure \ref{fig: figure3} (d)).

We repeat this procedure for other pump densities ($\dbath = 4.0$ and $6.3 \times10^{11}$ cm$^{-2}$) and extract values for the impurity-impurity interactions $\gii(\dbath)$ as function of the bath density, which we plot in Figure \ref{fig: figure4} (e) for the AP (red) and RP (blue) polaritons. For AP-polaritons the  interactions become more repulsive upon increasing the bath density. In contrast, for the RP-polaritons, the dressing by increasing the density of bath particles leads to a reduction of the polaron-interactions compared to the  bare repulsive polariton interactions. For the largest bath density that is experimentally achievable we even find that the interactions between RP-polaritons becomes attractive. This observation demonstrates that by polaron dressing one can not only strongly modify the magnitude of polariton interactions but even change their sign and turn repulsive interaction between bare impurity particles into net attractive interactions upon Bose polaron formation.

To theoretically investigate the interaction between polarons, we compute the expectation value of the Hamiltonian \eqref{Eq:Hamiltonian} with respect to a two-polaron state
\begin{align}
E^{2\text{Pol}}=\frac{\langle \text{L}|\hat a\hat a\mathcal{H}\hat a^{\dagger}\hat a^{\dagger}|\text{L}\rangle}{\langle \text{L}|\hat a\hat a\hat a^{\dagger}\hat a^{\dagger}|\text{L}\rangle}.
\label{eq:TwoPolarons}
\end{align}
In this first-order model the correlations between the polarons are neglected on the level of the wave function (up to the normalisation). The dependence of $E^{2\text{Pol}}$ on the probe exciton density is extracted by varying the system volume at fixed pump polariton density $\dbatht = N/V$.
From the dependence of the attractive and repulsive polaron energy on the ratio between probe and pump densities $\dimp/\dbath$, we  extract the effective interaction strength of the polaron-polaritons, which  is shown in Fig.~\ref{fig: figure4} (f) as a function of the bath density.  
Remarkably, already within the simple Ansatz~\eqref{eq:TwoPolarons} the 
same qualitative behaviour of the energy shift as the one observed in the experiment is obtained. However, the extracted effective interaction strengths differ by a factor 5 for the RPs, and by a factor 10 for the APs which we attribute to the uncertainty of the polariton density, the simplified Chevy-Ansatz based model used for the calculation of polaron-polaron interactions, as well as the modeling of the impurity density as arising from a simple finite size effect.

\textbf{Conclusion.}  
The experimental realisation of Bose polarons enables tunable polariton interactions in both magnitude and sign; this should be contrasted to polariton interactions in the absence of a bath  which are repulsive and weak. Our results present an important step towards the study of interactions in nonequilibrium polaronic systems which can guide further theoretical and experimental studies of novel quantum states of matter ranging from bipolarons \cite{Alexandrov_1992, Alexandrov_1994,Bruun2018a,Bruun2018b} to induced superconductivity \cite{Laussy2010,Cotlet16}.  For the case of bosonic, mobile impurities, our experiment demonstrates that polaron effects can turn repulsive interactions into attractive ones. Applying the observed Bose polaron effects to an electronic system may thus allow to weaken or even overcome the Coulomb repulsion between electrons \cite{Schmidt2022} opening up avenues for identifying new unconventional mechanisms of electron pairing in van der Waals materials based on exciton or polariton exchange.\\

{\large{\textbf{Methods}}}\\

\textbf{Experiment}\\
The experiment is performed on a semiconductor heterostructure in a zero-dimensional open fibre cavity.The empty cavity has a finesse of $\mathcal{F} \approx 6300$. The heterostructure  consists from top to bottom, of a hBN layer (32nm), a MoSe$_2$ layer, a hBN layer (90nm) and a graphene layer. The graphene and MoSe$_2$ are each contacted using metal electrodes made of Gold and Titanium. The individual layers were obtained by exfoliation on SiO$_2$/Si substrates and their thickness was obtained by AFM measurements. The van der Waals heterostructure was stacked by standard dry pickup technique and deposited on a fused Silica substrate with DBR-coating where the alternating layers are made up of Nb$_2$O$_5$ and SiO$_2$. The DBR coating was designed such that the node of the electric field coincides with the surface so that the bottom graphene gate does not lead to appreciable degradation of the cavity performance. 
The fibre core was ablated using CO$_2$ laser to form a dimple with 30 $\mu$m radius of curvature and coated with the same DBR structure as the substrate. The cavity device is mounted in a dipstick which is filled with low pressure Helium exchange gas, lowered into a liquid Helium bath and cooled to cryogenic temperatures for all measurements. The setup allows for optical access to the cavity either through the fibre, or from the substrate side through a free-space confocal microscope.

The time-resolved measurements are performed using a Ti:Sapphire mode-locked pulsed laser with an output power of up to 1.5 W and 140 fs pulse duration. The output pulses are split into two copies, the pump and the probe pulses respectively. The pump pulses are spectrally filtered to a FWHM of $\leq 0.5$ meV using a $4f$ pulse shaper while the probe pulses are pre-compensated for dispersion using a four-pass, free-space, single grating pulse compressor setup. The time delay $\tau$ between the pump and the probe pulses is controlled with a delay stage. The pump and the probe pulses are then sent through individual optical fibres to the dipstick for the measurements. At the sample position, the pump and probe pulses are measured to be 3 ps and 200 fs in duration respectively. The optical excitation is done via the free-space access to the cavity, which allows for individiually controlling their respective polarisation, while the transmission signal is collected from the fibre access of the cavity. The transmission signal is polarisation filtered to suppress the transmitted pump signal from reaching the detector. The signal is measured using a spectrometer with a grating of 1500 grooves per mm and a liquid Nitrogen cooled CCD camera.

In order to estimate the polariton density, we measure the incident pump power $P_\mathrm{pump}$ in the free-space path just before the entrance window into the dipstick. The effective pump power that excites the system is $\epsilon \eta P_\mathrm{pump}$ where $\epsilon$ is the spectral overlap of the pump and the polariton mode and $\eta$ is the spatial overlap of the pump mode and the cavity mode. $\epsilon$ is determined from the pump laser spectrum and the transmitted polariton spectrum while $\eta$ is determined from the signal reflected from the cavity upon white light excitation. The number of photons per pulse can be determined as $N_\mathrm{pulse} = \frac{\epsilon \eta P_\mathrm{pump}}{\hbar \omega f_\mathrm{rep}}$ where $f_\mathrm{rep} = 76$ MHz is the repetition rate of the pulsed laser. This corresponds to the integral under the temporal profile of the pump pulse. The time evolution of the polariton number in the system can then be obtained by convolving the exponential response function of the polariton with the pump temporal profile.
\\

\textbf{Theory}\\
We include a finite lifetime for the probe photon and the pump polariton by adding an imaginary part to the single particle energies in \eqref{Eq:Hamiltonian}, i.e.
$\omega_\mathrm{C}(\kv)\rightarrow \omega_\mathrm{C}(\kv)-i\gamma_\mathrm{C}$,
$\omega_\mathrm{LP}(\kv)\rightarrow \omega_\mathrm{LP}(\kv)-i\gamma_\mathrm{LP}$, with  $\gamma_\mathrm{C}$ the inverse lifetimes of the probe photon. The inverse lifetime of the pump polariton is given by $\gamma_\mathrm{LP}=-(1-\text{cos}^2\theta_{\mathbf{0}})\gamma_\mathrm{C}$ where 
\begin{align}
\text{cos}\,\theta_{\kv}=\frac{1}{\sqrt{2}}\sqrt{1+\frac{|\omega_\mathrm{C}(\kv)-\omega_\mathrm{X}(\kv)|}{\sqrt{(\omega_\mathrm{C}(\kv)-\omega_\mathrm{X}(\kv))^2+\Omega^2}}}
\end{align}
is the Hopfield factor. We do not consider a finite lifetime for the probe exciton in order to reduce the number of fit parameters.

The linear transmission probe spectrum is related 
to the impurity Green's function by $\mathcal{T} (\kv, E) = |\mathbf{G}_{22}(\kv, E)|^2$, which, in the exciton-photon basis, reads
\begin{align}
\mathbf{G}(E)=
    \begin{pmatrix}
               \omega_\mathrm{X}(0)
               -E+\Sigma(E) & \frac{\Omega}{2} \\
               \frac{\Omega}{2} & \omega_\mathrm{C}(0)-i\gamma_\mathrm{C}-E 
    \end{pmatrix}^{-1}.
\label{eq:Greensfunction}
\end{align}
Within our Ansatz, the self-energy takes the form $\Sigma(E)=\text{cos}^2\theta_{\mathbf{0}}\,n /(g^{-1}+V^{-1}\sum_{\kv}\text{cos}^2\theta_{\kv}/\xi_{\kv})$ with $\xi_{\kv}=\omega_\mathrm{X}(\kv)+(\omega_\mathrm{LP}(\kv)-\omega_\mathrm{LP}(0))-i\gamma_\mathrm{LP}-E$.
Since any purely attractive, short-range interaction supports a bound state in 2D, the interaction parameter $g$ can be related to the binding energy $\EXXb$ and inverse lifetime $\gamma_\mathrm{XX}$ of the biexciton via  $-1/g=V^{-1}\sum_{\kv}1/(\EXXb-i\gamma_\mathrm{XX}+2\omega_\mathrm{X}(\kv))$ .
For the spectrum shown in Fig.~\ref{fig: figure2} we used the experimental values
$\Omega=15\,\text{meV}$, $\gamma_\mathrm{C}=0.1\,\text{meV}$, $\EXXb=29.35\,\text{meV}$  and $\omega_\mathrm{LP}(0)-\omega_\mathrm{X}(0)=-15.5\,\text{meV}$, and obtained 
$\gamma_\mathrm{XX}=2.5\,\text{meV}$, and
$\dbath=2.5\times 10^{11}\text{cm}^{-2}$ from fitting the experimental data. 

For the calculation of the two-polaron energy $E^{2\text{Pol}}$, we evaluate \eqref{eq:TwoPolarons} using the variational parameters obtained from minimising the energy functional for the single polaron state.
Note that the Hamiltonian itself does not contain an interaction between impurities: the probe excitons exclusively interact through the presence of a bath. It is therefore crucial to work with a fixed particle number of pump polaritons in the Ansatz \eqref{Eq:Wavefunction}.\\

{\large{\textbf{Data availability}}}\\
The data are available at the ETH Research Collection (http://hdl.handle.net/20.500.11850/586847).\\

{\large{\textbf{Acknowledgements}}}\\
We thank Andrea Bergschneider for support in fabrication of the fibre dimple and design of the DBR coating. The work at ETH Zurich was supported by the Swiss National Science Foundation (SNSF) under Grant Number 200020\_207520. R.S. acknowledges support by the Deutsche Forschungsgemeinschaft under Germany’s Excellence Strategy - EXC 2181/1 - 390900948 (the Heidelberg STRUCTURES Excellence Cluster) and – EXC-2111 – 390814868. O. K. D. acknowledges funding from the International Max Planck Research School for Quantum Science and Technology (IMPRS - QST).\\

{\large{\textbf{Author contributions}}}\\
L.B.T. designed and built the experimental setup and performed the experiment with support of M.K.. L.B.T. and A.P. fabricated the sample. O.K.D. performed the theoretical calculations. L.B.T, O.K.D, R.S., A.I., and M.K. wrote the manuscript. R.S., A.I., and M.K. supervised the project.

\clearpage
\appendix
\setcounter{equation}{0}
\setcounter{figure}{0}
\setcounter{table}{0}
\setcounter{page}{1}
\makeatletter
\renewcommand{\theequation}{S\arabic{equation}}
\renewcommand{\thefigure}{S\arabic{figure}}

\bibliography{biblio}

\begin{figure*}[t!]
            \includegraphics[width=\textwidth]{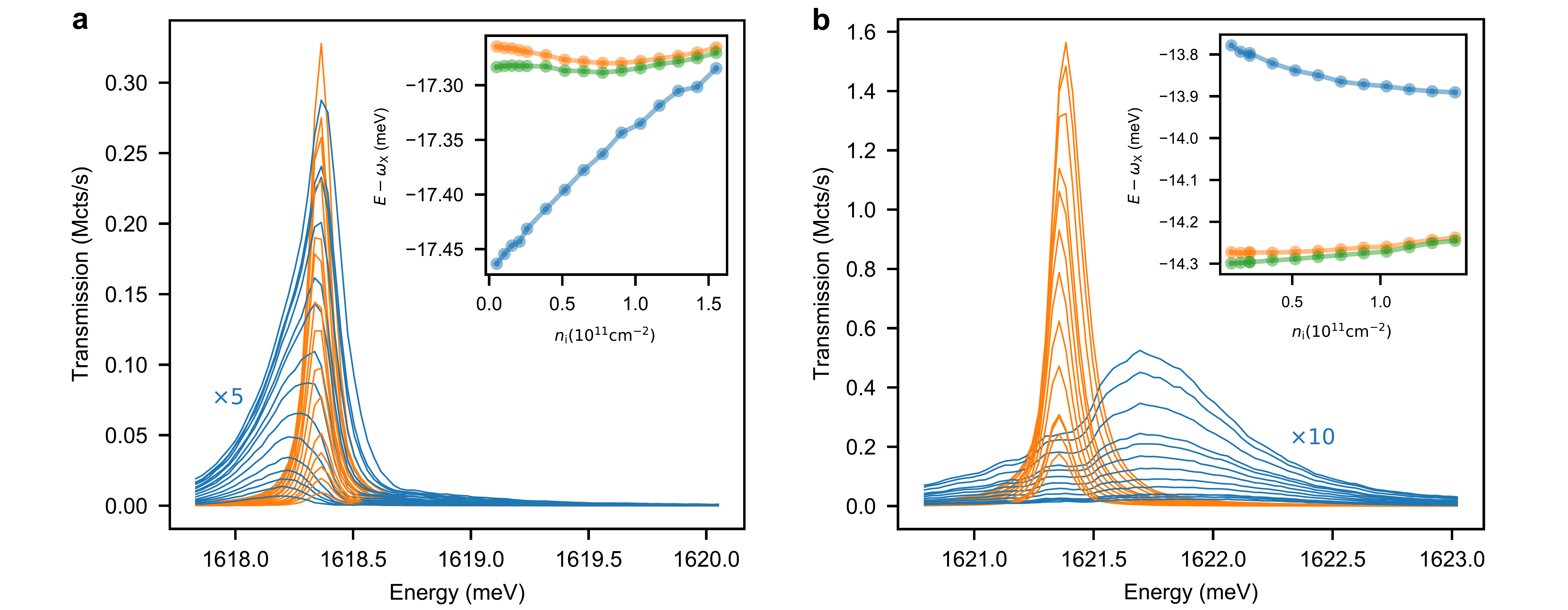}
    \caption{Impurity density dependent transmission spectra, as in Fig. \ref{fig: figure4} (a) and (b) but not normalised, for the attractive and repulsive polaron in (a) and (b) respectively. The blue lines correspond to the polaron spectrum at $\tau$ = 0 ps. The data was multiplied by the indicated factor for clarity. The orange lines represent the reference transmission spectra of the probe laser at $\tau$ = -50 ps. The insets show the extracted resonance energies of the probe laser transmission spectra as function of the impurity density $\dimp$. The blue dots were obtained from fitting the polaron spectrum (at $\tau$ = 0 ps), the orange dots from the reference spectrum ($\tau$ = -50 ps), and the green dots from the probe laser transmission spectrum with the pump laser blocked from reaching the sample. The resonance energies are plotted with respect to the exciton energy $\EX$.}
    \label{fig:EDF}
\end{figure*}

\end{document}